\documentclass[a4paper,11pt]{article}
\pdfoutput=1 

\usepackage{jheppub} 

\usepackage[T1]{fontenc} 
\usepackage{placeins}
\usepackage[dvipsnames,svgnames,x11names,hyperref]{xcolor}
\usepackage{array}
\usepackage{graphicx}
\usepackage{booktabs}
\usepackage{slashed}
\RequirePackage{mathrsfs}
\usepackage{dsfont}
\usepackage{hyperref}
\hypersetup{
        unicode = true,
        pdftitle = Impact of a XENONnT Signal on LHC Dijet Searches, 
        colorlinks = true,
        linkcolor = Black!30!DodgerBlue4,
        citecolor = Black!30!DodgerBlue4,
        filecolor = Black!30!DodgerBlue4,
        urlcolor = Black!30!DodgerBlue4,
}

\usepackage{booktabs}

\renewcommand{\toprule}{\midrule}
\renewcommand{\bottomrule}{\midrule}
\newcommand{\boldmidrule}{\midrule}

\graphicspath{{./img/}}

\newcommand{\unit}[1]{\,\mathrm{#1}}

\newcommand{\gev}{\unit{GeV}}
\newcommand{\tev}{\unit{TeV}}
\newcommand{\fb}{\unit{fb}}
\newcommand{\invfb}{\fb^{-1}}

\newcommand{\ii}{\mathrm{i}}
\newcommand{\ee}{\mathrm{e}}

\newcommand{\hc}{\mathrm{h.c.}}

\renewcommand{\epsilon}{\varepsilon}

\newcommand{\reffig}[1]{Fig.~\ref{#1}}

\renewcommand{\vec}[1]{\boldsymbol{#1}}
\renewcommand{\bar}{\overline}
\renewcommand{\Re}{\mathrm{Re}}
\renewcommand{\Im}{\mathrm{Im}}

\newcommand{\eg}{\textit{e.g.}}

\title{\boldmath Impact of a XENONnT Signal on LHC Dijet Searches}

\author[a,b]{Sebastian Baum}
\author[c]{Riccardo Catena}
\author[c,d]{and Martin B. Krauss}

\affiliation[a]{Oskar Klein Centre, Department of Physics, Stockholm University, AlbaNova, Stockholm SE-10691, Sweden}
\affiliation[b]{Nordita, KTH Royal Institute of Technology and Stockholm University, Roslagstullsbacken 23, 10691 Stockholm, Sweden}
\affiliation[c]{Chalmers University of Technology, Department of Physics, SE-412 96 G\"oteborg, Sweden}
\affiliation[d]{Dipartimento di Matematica e Fisica, Università di Roma Tre, Via della Vasca Navale 84, 00146 Rome, Italy}

\emailAdd{sbaum@fysik.su.se}
\emailAdd{catena@chalmers.se}
\emailAdd{martin.krauss@chalmers.se}

\abstract{It is well-known that dark matter (DM) direct detection experiments and the LHC are complementary, since they probe physical processes occurring at different energy scales.~And yet, there are aspects of this complementarity which are still not fully understood, or exploited.~For example, what is the impact that the discovery of DM at XENONnT would have on present and future searches for DM in LHC final states involving a pair of hadronic jets?~In this work we investigate the impact of a XENONnT signal on the interpretation of current dijet searches at the LHC, and on the prospects for dijet signal discovery at the High-Luminosity (HL) LHC in the framework of simplified models.~Specifically, we focus on a general class of simplified models where DM can have spin 0, 1/2 or 1, and interacts with quarks through the exchange of a scalar, pseudo-scalar, vector, or pseudo-vector mediator.~We find that exclusion limits on the mediator's mass and its coupling to quarks from dijet searches at the LHC are significantly affected by a signal at XENONnT, and that $\mathcal{O}(100)$ signal events at XENONnT would drastically narrow the region in the parameter space of simplified models where a dijet signal can be discovered at $5\sigma$ C.L. at the HL-LHC.} 

\begin{document} 
\maketitle
\flushbottom

\section{Introduction}
Observations in a wide range of astronomical and cosmological systems show that the Universe contains about five times as much dark matter (DM) as baryonic matter~\cite{Bertone:2004pz}.~While the nature of DM remains unknown, the hypothesis that DM is made of yet unidentified particles is the one explored most extensively~\cite{Bertone:2016nfn}.~If DM is made of Weakly Interacting Massive Particles (WIMPs), as  predicted by many theories beyond the Standard Model (BSM) addressing the hierarchy problem, e.g. supersymmetric extension of the Standard Model (SM), it can potentially be observed in the next stage of direct detection and particle collider experiments~\cite{Arcadi:2017kky}.~DM direct detection experiments primarily search for nuclear recoils induced by the non-relativistic scattering of Milky Way DM particles in low-background detectors located deep underground~\cite{Drukier:1983gj,Goodman:1984dc}.~The null results of present DM direct detection experiments place severe constraints on the strength with which DM couples to the fundamental constituents of matter.~The most stringent limits from direct detection experiments on the strength of DM-nucleus interactions for WIMPs heavier than about 10 GeV are currently set by the  XENON1T collaboration~\cite{Aprile:2018dbl}, improving on previous results from LUX~\cite{daSilva:2017swg} and PandaX-II~\cite{Cui:2017nnn}.~They reported  a 90\% C.L. exclusion limit on the elastic spin-independent DM-nucleon scattering cross-section with a minimum of  $4.1\times10^{-47}\unit{cm^2}$ around a DM mass of $30\gev$.~XENONnT, the upgrade of XENON1T, is expected to operate from 2019 onwards using about 7 ton of ultra-pure liquid xenon as target material~\cite{aprile:2015uzo}. Note that although we use XENONnT as an example for the purposes of this work, the LZ~\cite{Mount:2017qzi} and PandaX-4T~\cite{Zhang:2018xdp} experiments plan to achieve similar sensitivity on comparable time scales as XENONnT. Pursuing a complementary approach to direct detection experiments, the Large Hadron Collider (LHC) at CERN with data analysed by the ATLAS and CMS collaborations is also searching for DM.~At the LHC, DM can be produced in the collision of energetic protons, and its production be inferred through the observation of missing transverse momentum in the final state of such collisions.~The next run of the LHC (Run 3) will start in 2021~\cite{lhc-schedule}.~It will operate at the centre-of-mass energy of $\sqrt{s}=13$~TeV and reach the expected integrated luminosity of 300 fb$^{-1}$ in 2023.~The LHC Run 3 will be followed by the high luminosity run of the LHC (HL-LHC), which is expected to start in 2026, reaching an integrated luminosity of  3000 fb$^{-1}$.

LHC data on processes that might involve DM have been interpreted within different theoretical frameworks.~Ultraviolet (UV) complete theories (with a focus on supersymmetric theories) and Effective Field Theories (EFTs) for DM-quark and -gluon interactions have been used extensively in the analyses of the LHC Run 1 results (see~\cite{DeSimone:2016fbz}, and references therein).~Limitations in the applicability of an EFT approach to the interpretation of LHC data led to a change of framework for the interpretation of results from the LHC Run 2~\cite{Buchmueller:2013dya}, which were primarily analysed within the framework of simplified models (see ~\cite{Abdallah:2015ter}, and references therein).~By construction, in simplified models for DM, the SM is extended by the DM particle and one single mediator particle only.~The latter is responsible for the interactions of DM with SM particles, for example, quarks and gluons.~When the momentum transferred in a proton-proton collision at the LHC is smaller than the mediator mass (which is not a priori true), the mediator can be ``integrated out'' and simplified models converge to EFTs.~Compared to an EFT approach, the use of simplified models allows for a more complex analysis of the LHC data, especially in the study of processes which involve the mediator explicitly, as in the case of final states including hadronic jets produced by the decay of the mediator into a quark pair.~While simplified models are generically not UV complete, and their applicability is subject to constraints from unitarity and anomaly cancellation, see e.g.~\cite{Kahlhoefer:2015bea,Ellis:2017tkh}, they provide a good compromise between simplicity and completeness.

While processes that directly involve the DM particle are obviously important to reconstruct DM mass and coupling constants at the LHC and have therefore been studied, \eg, in monojet searches~\cite{Sirunyan:2017jix,Aaboud:2017phn}, events involving the mediator particle alone can be used to obtain important information on the underlying DM model as well~\cite{Chala:2015ama}.~In this context, dijet searches play a special role~\cite{Abercrombie:2015wmb}.~Within the framework of simplified models for DM-quark interactions, neutral mediators can be resonantly produced during a proton--proton collision and then decay into a pair of quarks.~Due to hadronization these will be seen by a detector as a pair of hadronic jets.~The analysis of dijet events at ATLAS and CMS has been one of the main channels in the search for new physics at the LHC.~Recently, ATLAS has published results for generic (high-mass) dijet searches from $37\invfb$ of data collected during 2015 and 2016~\cite{Aaboud:2017yvp}, whereas CMS has presented results for $36\invfb$ of data from the 2016 dataset~\cite{CMS-PAS-EXO-16-056,Sirunyan:2018xlo}, as well as preliminary results for $78\invfb$ of data from the combined 2016 and 2017 datasets~\cite{CMS-PAS-EXO-17-026}.\footnote{ATLAS has published data including the 2017 data set only for a dijet search requiring an additional isolated lepton in the final state~\cite{ATLAS-CONF-2018-015}.}~In these studies, both collaborations have presented results only for a subset of all possible simplified models, mainly focusing on a vector mediator and fermionic DM.

Direct detection experiments and the LHC are complementary since they explore 
physical processes occurring at different energy scales, see 
e.g.~Ref.~\cite{Baltz:2006fm}.~Nevertheless, there are aspects of this 
complementarity which are still not fully understood, or exploited.~For example, 
what is the impact of a signal at XENONnT on present and future searches for DM 
and new physics in general at the LHC?~In this work we investigate the impact of 
a XENONnT signal on the interpretation of current dijet searches at the LHC, and 
on the prospects for dijet signal discovery at the HL-LHC.~As a theoretical 
framework, we use a general class of simplified models where DM can have spin 0, 
1/2 or 1, and interacts with quarks through the exchange of scalar, 
pseudo-scalar, vector, or pseudo-vector mediators.~This study extends our 
previous work~\cite{Baum:2017kfa}, where we focused on the impact of a XENONnT 
signal on monojet searches at the LHC.~We find that exclusion limits on mediator 
parameters from dijet searches at the LHC are significantly affected by a signal 
at XENONnT, and that $\mathcal{O}(100)$ signal events at XENONnT dramatically 
narrow the region in the parameter space of simplified models where a dijet 
signal can be discovered at $5\sigma$ C.L.

We will show that there is an interesting interplay at work between direct detection and dijet searches, since the mass of the mediator and its couplings to quarks and DM are correlated when a direct detection signal is observed. If one fixes the mediator mass and one coupling, the other coupling will be fully determined by the direct detection signal. Since direct detection is sensitive to the product of the couplings of the mediator to quarks and to DM only, it is possible that the dijet signal is too small to be detectable if the mediator--DM coupling is large and the mediator--quark coupling small. However, we will show that sizable regions of parameter space remain, where both a signal in direct detection and in dijet searches at the LHC can potentially be detected. Furthermore, the relation between direct detection and dijet signals is model dependent. Thus, the combination of the two approaches allows one to gain additional information about the model and to break degeneracies between different models which appear when considering either direct detection or dijet searches only.

The remaining part of this article is organised as follows.~In Secs.~\ref{sec:theory} and~\ref{sec:stat} we review theoretical framework and statistical methods.~Our analysis of the impact of a XENONnT signal on LHC dijet searches is illustrated in Sec.~\ref{sec:method}.~We conclude in Sec.~\ref{sec:conclusion} and list useful equations in Appendix~\ref{sec:app}. 

\section{Theoretical framework}
\label{sec:theory}
The theoretical framework used in this work consists of a set of simplified 
models where DM can have spin 0, 1/2 or 1, and interacts with quarks through the 
exchange of scalar, pseudo-scalar, vector or pseudo-vector 
mediators~\cite{dent:2015zpa}. 
Beyond the spin of the DM candidate and the mediator, the simplified models are specified by the Lorentz structure of the corresponding interaction vertices, and four free parameters.~For each simplified model, the four free parameters are the DM particle and mediator masses, $m_{\rm DM}$ and $m_{\rm med}$, respectively, and two coupling constants:~one for the DM-DM-mediator vertex, $g_{\rm DM}$, and the second one for the quark-quark-mediator vertex, $g_{q}$.\footnote{Note that in particular all relevant quantities for dijet searches, i.e. the mediator's production cross section, its width, and its branching ratios, are fixed by these four parameters.}~Presenting our results, we specifically focus on nine simplified models characterised by the interaction Lagrangians 
\begin{align}
\mathcal{L}_1 &=-h_1\bar{q} q\phi -g_1m_SS^\dagger S\phi \,;& \quad &(S\otimes S)_0 \nonumber \\
\mathcal{L}_2 &=-h_3(\bar{q}\gamma_{\mu}q)G^{\mu}-ig_4(S^{\dagger}\partial_{\mu}S-\partial_{\mu}S^{\dagger}S)G^{\mu}\,;& \quad &(V\otimes i\partial)_0 \nonumber \\
\mathcal{L}_3 &=-h_1\phi\bar{q} q-\lambda_1\phi\bar{\chi}\chi \,;& \quad &(S\otimes S)_{1/2} \nonumber \\
\mathcal{L}_4 &=-h_1\phi\bar{q} q-i\lambda_2\phi\bar{\chi}\gamma^{5}\chi\,;& \quad &(S\otimes PS)_{1/2} \nonumber \\
\mathcal{L}_5 &=-h_3\bar{q}\gamma_{\mu}qG^{\mu}-\lambda_{3}\bar\chi\gamma^\mu\chi G_{\mu} \,;& \quad &(V\otimes V)_{1/2} \nonumber \\
\mathcal{L}_6 &=-h_3\bar{q}\gamma_{\mu}qG^{\mu}-\lambda_{4}\bar\chi\gamma^\mu\gamma^5\chi G_{\mu}  \,;& \quad &(V\otimes A)_{1/2} \nonumber \\
\mathcal{L}_7 &=-h_4\bar{q}\gamma_{\mu}\gamma^{5}qG^{\mu}-\lambda_{4}\bar\chi\gamma^\mu\gamma^5\chi G_{\mu}  \,; &\quad &(A\otimes A)_{1/2} \nonumber \\
\mathcal{L}_8 &=-h_1\phi\bar{q}q-b_1m_X\phi X_{\mu}^{\dagger}X^{\mu} \,;& \quad &(S\otimes S)_1 \nonumber \\
\mathcal{L}_9 &=-h_3G_\mu\bar{q}\gamma^\mu q-ib_{5}(X_{\nu}^{\dagger}\partial_{\mu}X^{\nu}-X^\nu \partial_\mu X^\dagger_\nu)G^\mu\,;& \quad &(V\otimes i\partial)_1
\label{eq:simpL}
\end{align}
where $g_q=h_1, h_2, h_3$ or $h_4$ and $g_{\rm DM}= \lambda_1, \lambda_2, 
\lambda_3, \lambda_4, g_1, g_4, b_1$ or $b_5$, depending on the model.~In 
Eq.~(\ref{eq:simpL}), scalar, fermionic and vector DM are described by the 
complex scalar field $S$, the spinor field $\chi$ and the complex vector field 
$X_\nu$, respectively.~Quarks spinors are denoted by $q$, and a summation on 
quark flavours is understood.~Next to each interaction Lagrangian, we have 
introduced a ``label'' which will be used in the following to refer to the 
corresponding simplified model.~For example, $(S\otimes S)_0$, is the label for 
spin 0 DM coupling to quarks via a scalar DM-DM-mediator vertex and a scalar 
quark-quark-mediator vertex.~Similarly, $(V\otimes i\partial)_1$ refers to spin 
1 DM coupling to quarks via a vector quark-quark-mediator vertex and a 
derivative DM-DM-mediator vertex.~In the remaining cases, the letter $A$ refers 
to axial coupling and $PS$ to pseudo-scalar coupling.~As we will explain later, 
the simplified models in Eq.~(\ref{eq:simpL}) form the subset of models 
from~\cite{dent:2015zpa} that are compatible with the discovery of 
$\mathcal{O}(100)$ signal events at XENONnT and the current search for narrow 
resonances in dijet final states at the LHC.~For completeness, the full set of 
simplified models from~\cite{dent:2015zpa} is reported in 
Appendix~\ref{sec:app}.~These models were introduced in the context of DM direct 
detection~\cite{dent:2015zpa}, and later applied to LHC monojet 
analyses~\cite{Baum:2017kfa} and DM relic density 
calculations~\cite{Catena:2017xqq}.

For each simplified model in Eq. (\ref{eq:simpL}), we simulate dijet signals at the LHC, and calculate the corresponding dijet invariant mass spectrum, by using the chain of numerical programs 
\\
\begin{minipage}[c]{\linewidth}
\begin{center}
\texttt{WHIZARD \\
 $\Downarrow$ \\
 pythia8 \\
 $\Downarrow$ \\
 Delphes3 (+FASTJET) \\ 
$\Downarrow$ \\
Custom C++/ROOT code for analysis.
}
\end{center}
\end{minipage}
\\
\\
We use \texttt{WHIZARD}~\cite{Kilian:2007gr,Moretti:2001zz} with model files implementing the simplified models to generate the hard processes
\begin{equation}
 p, p \rightarrow {\rm Mediator} \rightarrow \bar q q + X
\end{equation}
where $X$ stands for additional SM particles and $\bar q q$ can be any pair of 
quarks with the same flavor.~We use parton distribution functions  from the 
\texttt{CT14lo} set as obtained from 
\texttt{LHAPDF6}~\cite{Buckley:2014ana},~\texttt{pythia8}~\cite{
Sjostrand:2007gs} is used for showering and hadronization, 
\texttt{Delphes3}~\cite{deFavereau:2013fsa} for (CMS) detector simulation, and 
\texttt{FASTJET}~\cite{Cacciari:2011ma} for jet reconstruction.~We use our own 
\texttt{C++} code and \texttt{ROOT}~\cite{Brun:1997pa} libraries to analyse the 
signal.~We discard events where one (or both) of the leading jets deposit more 
than 90\% of their respective total calorimetric energy in the electromagnetic 
calorimeter.~See~\cite{Baum:2018sxd} for a more detailed discussion of our 
collider simulations.

\begin{table}[t]
\begin{center}
\begin{tabular}{lcl}
\toprule
  Spin 0 DM & $\qquad\qquad\qquad\qquad$ &  Non-relativistic coefficients different from zero\\ \addlinespace[.4em]
  $(S\otimes S)_0$ & & $c_1^{(N)}=\frac{h_1^N g_1}{M^2_\Phi}$ \\  \addlinespace[.4em]
   $(V\otimes i\partial)_0$ & & $c_1^{(N)}=-2 \frac{h_3^N g_4}  { M_G^2}$   \\ \addlinespace[.4em]
    \boldmidrule
    spin 1/2 DM & $\qquad\qquad\qquad\qquad$ & Non-relativistic coefficients different from zero\\ \addlinespace[.4em]
   $(S\otimes S)_{1/2}$ & & $c_1^{(N)}=\frac{h_1^N \lambda_1}{M^2_\Phi}$ \\ \addlinespace[.4em]  
   $(S\otimes PS)_{1/2}$ & & $c_{11}^{(N)}=-\frac{h_1^N \lambda_2}{M_\Phi^2}\frac{m_N}{m_\chi}$\\ \addlinespace[.4em]
   $(V\otimes V)_{1/2}$ & & $c_1^{(N)}=-\frac{h_3^N \lambda_3}{M_G^2}$\\ \addlinespace[.4em]
   $(V\otimes A)_{1/2}$ & &  $c_{8}^{(N)}=-2\frac{h_3^N \lambda_4}{M_G^2}\,; \quad c_{9}^{(N)}=-2\frac{h_3^N \lambda_4}{M_G^2}$ \\ \addlinespace[.4em]
   $(A\otimes A)_{1/2}$ & & $c_4^{(N)}=4\frac{h_4^N \lambda_4}{M_G^2}$ \\ \addlinespace[.4em]
  \boldmidrule
    spin 1 DM & $\qquad\qquad\qquad\qquad$ & Non-relativistic coefficients different from zero\\ \addlinespace[.4em]   
    $(S\otimes S)_1$ & &  $c_1^{(N)}=\frac{b_1 h_1^N}{M^2_\Phi}$ \\ \addlinespace[.4em]  
    $(V\otimes i\partial)_1$ & &   $c_1^{(N)}=-2 \frac{h_3^N b_5}  { M_G^2}$ \\ \addlinespace[.4em]  
     \bottomrule
 \end{tabular}  
\end{center}
    \caption{Relation between the coupling constants of non-relativistic operators from Tab.~\ref{tab:operators} (in the proton/neutron basis) and simplified models in this study (see Eq. (\ref{eq:simpL}) and Appendix~\ref{sec:app} for their Lagrangians).~In the case of spin 1 DM, we do not consider non-relativistic operators that depend on the symmetric combination of polarisation vectors denoted by $\mathcal{S}$ in Tab.~\ref{tab:operators}.}
    \label{tab:coeffs}
\end{table}

\begin{table}[t]
    \centering
    \begin{tabular*}{\columnwidth}{@{\extracolsep{\fill}}llll@{}}
    \toprule
        $\hat{\mathcal{O}}_1 = \mathds{1}_{\chi}\mathds{1}_N$  & $\hat{\mathcal{O}}_{10} = i{\bf{\hat{S}}}_N\cdot\frac{{\bf{\hat{q}}}}{m_N}\mathds{1}_\chi$   \\
        $\hat{\mathcal{O}}_3 = i{\bf{\hat{S}}}_N\cdot\left(\frac{{\bf{\hat{q}}}}{m_N}\times{\bf{\hat{v}}}^{\perp}\right)\mathds{1}_\chi$ & $\hat{\mathcal{O}}_{11} = i{\bf{\hat{S}}}_\chi\cdot\frac{{\bf{\hat{q}}}}{m_N}\mathds{1}_N$   \\
        $\hat{\mathcal{O}}_4 = {\bf{\hat{S}}}_{\chi}\cdot {\bf{\hat{S}}}_{N}$ & $\hat{\mathcal{O}}_{12} = {\bf{\hat{S}}}_{\chi}\cdot \left({\bf{\hat{S}}}_{N} \times{\bf{\hat{v}}}^{\perp} \right)$  \\
        $\hat{\mathcal{O}}_5 = i{\bf{\hat{S}}}_\chi\cdot\left(\frac{{\bf{\hat{q}}}}{m_N}\times{\bf{\hat{v}}}^{\perp}\right)\mathds{1}_N$ & $\hat{\mathcal{O}}_{13} =i \left({\bf{\hat{S}}}_{\chi}\cdot {\bf{\hat{v}}}^{\perp}\right)\left({\bf{\hat{S}}}_{N}\cdot \frac{{\bf{\hat{q}}}}{m_N}\right)$\\
        $\hat{\mathcal{O}}_6 = \left({\bf{\hat{S}}}_\chi\cdot\frac{{\bf{\hat{q}}}}{m_N}\right) \left({\bf{\hat{S}}}_N\cdot\frac{\hat{{\bf{q}}}}{m_N}\right)$ &  $\hat{\mathcal{O}}_{14} = i\left({\bf{\hat{S}}}_{\chi}\cdot \frac{{\bf{\hat{q}}}}{m_N}\right)\left({\bf{\hat{S}}}_{N}\cdot {\bf{\hat{v}}}^{\perp}\right)$  \\
        $\hat{\mathcal{O}}_7 = {\bf{\hat{S}}}_{N}\cdot {\bf{\hat{v}}}^{\perp}\mathds{1}_\chi$ &  $\hat{\mathcal{O}}_{15} = -\left({\bf{\hat{S}}}_{\chi}\cdot \frac{{\bf{\hat{q}}}}{m_N}\right)\left[ \left({\bf{\hat{S}}}_{N}\times {\bf{\hat{v}}}^{\perp} \right) \cdot \frac{{\bf{\hat{q}}}}{m_N}\right] $ \\
        $\hat{\mathcal{O}}_8 = {\bf{\hat{S}}}_{\chi}\cdot {\bf{\hat{v}}}^{\perp}\mathds{1}_N$ &  $\hat{\mathcal{O}}_{17}=i \frac{{\bf{\hat{q}}}}{m_N} \cdot \boldsymbol{\mathcal{S}} \cdot {\bf{\hat{v}}}^{\perp} \mathds{1}_N$ \\
        $\hat{\mathcal{O}}_9 = i{\bf{\hat{S}}}_\chi\cdot\left({\bf{\hat{S}}}_N\times\frac{{\bf{\hat{q}}}}{m_N}\right)$ & $\hat{\mathcal{O}}_{18}=i \frac{{\bf{\hat{q}}}}{m_N} \cdot \boldsymbol{\mathcal{S}}  \cdot {\bf{\hat{S}}}_{N}$  \\
    \hline
    \end{tabular*}
    \caption{Quantum mechanical operators defining the non-relativistic 
effective theory of DM-nucleon interactions~\cite{Fitzpatrick:2012ix}.~The 
operators are expressed in terms of the basic invariants under Galilean 
transformations:~the momentum transfer, $\boldsymbol{\hat{q}}$, the transverse 
relative velocity operator $\boldsymbol{\hat{v}}^\perp$, the nucleon and DM spin 
operators, denoted by $\boldsymbol{\hat{S}}_N$ and $\boldsymbol{\hat{S}}_\chi$, 
respectively, and the identities in the nucleon and DM spin spaces, 
$\mathds{1}_{\chi}$ and $\mathds{1}_N$.~All operators have the same mass 
dimension, and $m_N$ is the nucleon mass.~Standard spin-independent and 
spin-dependent interactions correspond to the operators $\hat{\mathcal{O}}_{1}$ 
and $\hat{\mathcal{O}}_{4}$, respectively, while $\boldsymbol{\mathcal{S}}$ is a 
symmetric combination of spin 1 polarisation vectors~\cite{dent:2015zpa}.~The 
operators $\hat{\mathcal{O}}_{17}$ and $\hat{\mathcal{O}}_{18}$ can only arise 
for spin 1 DM.~Following~\cite{Fitzpatrick:2012ix}, here we do not consider the 
interaction operators $\hat{\mathcal{O}}_{2}$ and $\hat{\mathcal{O}}_{16}$:~the 
former is quadratic in $\boldsymbol{\hat{v}}^\perp$ (and the effective theory 
expansion in~\cite{Fitzpatrick:2012ix} is truncated at linear order in 
$\boldsymbol{\hat{v}}^\perp$ and second order in $\boldsymbol{\hat{q}}$) and the 
latter is a linear combination of $\hat{\mathcal{O}}_{12}$ and 
$\hat{\mathcal{O}}_{15}$.}
\label{tab:operators}
\end{table}

For each simplified model in Eq. (\ref{eq:simpL}), we are also interested in the rate of DM-nucleus scattering events at XENONnT.~The expected rate per unit detector mass can be written as
\begin{align}
\frac{{\rm d} R}{{\rm d} E_R} = \sum_T \xi_T \frac{\rho_\chi}{m_\chi m_T} \int_{|\mathbf{v}| \ge v_{\rm min}} {\rm d}^3 v \, |\mathbf{v}| f(\mathbf{v}) \, \frac{{\rm d}\sigma_T}{{\rm d} E_R} (|\mathbf{v}|^2, E_R) \,,
\label{eq:rate}
\end{align}
where $v_{\rm min}$ is the minimum DM speed to deposit an energy $E_R$ in the 
detector, ${\rm d}\sigma_T/{\rm d} E_R$ is the differential cross section for 
DM-nucleus scattering, $\rho_\chi$ is the local DM density, and $f(\mathbf{v})$ 
is the local DM velocity distribution in the detector rest frame.~In the sum in 
Eq.~(\ref{eq:rate}), we consider the seven most abundant xenon isotopes.~Their 
mass fraction is denoted here by $\xi_T$.~In order to calculate the expected 
rate of DM-nucleus scattering events at XENONnT for the simplified models in Eq. 
(\ref{eq:simpL}), we proceed as follows.~First, we analytically calculate the 
amplitude for DM scattering on free nucleons as described in detail 
in~\cite{Catena:2018uae,Baum:2017kfa,Bishara:2017nnn,Catena:2015uha,
Catena:2014epa,Catena:2014uqa,DelNobile:2013sia}.~From these amplitudes, we 
extract the coupling constants for DM-nucleon interactions, which are related to 
the ones in the Lagrangians in Eq. (\ref{eq:simpL}), as illustrated in 
Tab.~\ref{tab:coeffs}.
~We then use the coupling constants in Tab.~\ref{tab:coeffs} as an input for the 
package \texttt{DMFormFactor}~\cite{anand:2013yka},
which provides a generalized set of nuclear response functions that properly 
treat velocity dependent WIMP interactions, and from which we extract the 
rate of DM-nucleus scattering events at XENONnT as an output.
~The result of 
this calculation depends on the local DM density and velocity distribution.~For 
the DM velocity distribution in the detector rest frame, we assume a Maxwellian 
velocity distribution with a circular speed of 220 km~s$^{-1}$ for the local 
standard of rest, and a galactic escape velocity of 544 km~s$^{-1}$.~Finally, 
for the local DM density we adopt the value 0.4~GeV~cm$^{-3}$.

The coupling constants in Tab.~\ref{tab:coeffs} are the coefficients of quantum mechanical operators defining the non-relativistic effective theory of DM-nucleon interactions~\cite{Dobrescu:2006au,Fan:2010gt,Fitzpatrick:2012ix}.~In this framework, DM-nucleon interaction operators are denoted by $\hat{\mathcal{O}}_{i}^{(N)}$, or just $\hat{\mathcal{O}}_{i}$ for simplicity.~We list them in Tab.~\ref{tab:operators} for completeness.~They are expressed in terms of the basic invariants under Galilean transformations and Hermitian conjugation, namely:~the momentum transfer operator, $\boldsymbol{\hat{q}}$, the transverse relative velocity operator $\boldsymbol{\hat{v}}^\perp$, the nucleon and DM spin operators, $\boldsymbol{\hat{S}}_N$ and $\boldsymbol{\hat{S}}_\chi$, respectively, and the identities in the nucleon and DM spin spaces, $\mathds{1}_{\chi}$ and $\mathds{1}_N$.~In Tab.~\ref{tab:operators}, $m_N$ is the nucleon mass, and all interaction operators have the same mass dimension.~Within this notation, standard spin-independent and spin-dependent interactions correspond to the operators $\hat{\mathcal{O}}_{1}$ and $\hat{\mathcal{O}}_{4}$, respectively.~The operators $\hat{\mathcal{O}}_{2}$ and $\hat{\mathcal{O}}_{16}$ do not appear in Tab.~\ref{tab:operators} for the following reasons:~the former is quadratic in $\boldsymbol{\hat{v}}^\perp$ (while the effective theory expansion in~\cite{Fitzpatrick:2012ix} is truncated at second order in $\boldsymbol{\hat{q}}$ and at linear order in $\boldsymbol{\hat{v}}^\perp$) and the latter is not independent, being a linear combination of the interaction operators $\hat{\mathcal{O}}_{12}$ and $\hat{\mathcal{O}}_{15}$.~Finally, the operators $\hat{\mathcal{O}}_{17}$ and $\hat{\mathcal{O}}_{18}$ in Tab.~\ref{tab:operators} can only arise for spin 1 DM, and $\boldsymbol{\mathcal{S}}$ is a symmetric combination of spin 1 polarisation vectors~\cite{dent:2015zpa}.~In terms of the interaction operators $\hat{\mathcal{O}}_{i}$ in Tab.~\ref{tab:operators}, each simplified model in Eq. (\ref{eq:simpL}) generates an Hamiltonian for non-relativistic DM-nucleon interactions, $\mathscr{H}$, which can be expressed as follows
\begin{equation}
\mathscr{H}= \sum_{N=p,n}\sum_{i} c_i^{(N)}\hat{\mathcal{O}}_{i}^{(N)} 
\,,
\label{eq:H}
\end{equation}
where $c_i^{(p)}$ and $c_i^{(n)}$ are the coupling constants for protons and neutrons, respectively.~For example, the simplified model characterised by fermionic DM and a vector mediator of mass $M_G$ that couples to DM with coupling constant $g_{\rm DM}=\lambda_3$ and to quarks with (a universal) coupling constant $g_q=h_3$ generates the operator $\hat{\mathcal{O}}_1$ in the non-relativistic limit.~In this case, $c_1^{(N)} = -h_3^N \lambda_3/M_G^2$, where the nucleon-level and quark-level coupling constants, $h_3^N$ and $h_3$, are related by $h_3^N=3 h_3$.~For expressions relating $h_i^{(N)}$ to $h_i$, with $i=1,2,4$, we refer to~\cite{dent:2015zpa}.~In the non-relativistic limit, some of the simplified models in Eq. (\ref{eq:simpL}) generate a linear combination of operators in~Tab.~\ref{tab:operators} (see Tab.~\ref{tab:coeffs}).~However, for $m_{\rm DM}=50$~GeV (the benchmark value used in our calculation), we find that it is always possible to identify a leading operator among those generated from a given simplified model in the non-relativistic limit.

\section{Statistical methods}
\label{sec:stat}
We compute exclusion limits and discovery regions (or sensitivity projections) using the profile likelihood ratio method~\cite{cowan:2010js}.~In the former case, we compare the background plus signal hypothesis, $H_1$, with the background only hypothesis, $H_0$, computing the significance with which a point in parameter space can be excluded.~In the latter case, we test the null hypothesis $H_0$ against the alternative $H_1$, computing the significance with which a point in parameter space can be observed.~In both cases, we obtain the significance, $Z$, from a profile likelihood ratio $\lambda$ and the test statistic $q = -2 \ln \lambda$, using standard asymptotic formulae from~\cite{cowan:2010js}
\begin{equation}
Z\simeq\sqrt{q} \,.
\end{equation}
The significance is also related to the $p$-value, i.e.~$Z = \Phi^{-1}(1-p)$, where $\Phi^{-1}$ is the quantile of a Gaussian distribution with mean 0 and variance 1.~For example, standard 95\% confidence level (C.L.) exclusion limits correspond to a $p$-value of $0.05$ and a significance of $1.64$.

The exact form of the profile likelihood ratio depends on whether we calculate discovery regions or exclusion limits.~For exclusion limits, the profile likelihood ratio takes the following form
\begin{equation} \label{eq:profL}
\lambda  =  \frac{\mathscr{L}(\boldsymbol{s},\widehat{\boldsymbol{\theta}})}{\mathscr{L}(\boldsymbol{0},\widehat{\widehat{\boldsymbol{\theta}}})} \,,
\end{equation}
where the likelihood function, $\mathscr{L}$, is defined below.~For discovery regions, the profile likelihood ratio is given by
\begin{equation} \label{eq:profL2}
\lambda  =  \frac{\mathscr{L}(\boldsymbol{0},\widehat{\widehat{\boldsymbol{\theta}}})}{\mathscr{L}(\boldsymbol{s},\widehat{\boldsymbol{\theta}})} \,.
\end{equation}
Here, the likelihood function for finding a given signal $\boldsymbol{s}=\{s_1,\dots,s_N\}$ over a background $\boldsymbol{b}=\{b_1,\dots,b_N\}$ for a dataset $\vec{n}=(n_1,\dots,n_N)$ is defined as the product of $N$ Poisson distributions 
\begin{equation}
  \mathscr{L}\left(\boldsymbol{s},\boldsymbol{\theta}\right) = \prod_{i=1}^N \frac{(s_i + 
b_i(\vec{\theta}))^{n_i}}{n_i!}\ee^{-[s_i + b_i(\vec{\theta})]}\,,
\end{equation}
where $N$ is the number of bins in the dijet invariant mass, $s_i$ the number of 
signal events in the $i$-th bin, $b_i$ the number of background events in the 
same bin, and $\boldsymbol{\theta}$ a set of nuisance parameters, i.e.~the 
background model parameters from \cite{1676214}, in our case.\footnote{For each 
parameter point independently, we maximise the likelihood over all invariant 
mass bins for a given $n_i$ (excluding a window around the mediator mass). The 
obtained model parameters include the systematic uncertainties on the 
background. The statistical uncertainties are accounted for by assuming a 
Poisson distribution of the events bin-by-bin around their expectation value.}
~In the definition(s) 
of $\lambda$, $\widehat{\boldsymbol{\theta}}$ 
($\widehat{\widehat{\boldsymbol{\theta}}}$) is the set of nuisance parameters 
maximizing the likelihood function for the given signal $\boldsymbol{s}$ 
($\boldsymbol{0}$).~Maximising $\mathscr{L}$ with respect to 
$\boldsymbol{\theta}$ to find $\widehat{\boldsymbol{\theta}}$ or 
$\widehat{\widehat{\boldsymbol{\theta}}}$ at each point in parameter space, we 
exclude a window around the mediator mass in the dijet invariant mass 
spectrum.~See~\cite{Baum:2018sxd} for further details.

Computing exclusion limits, we evaluate $Z$ for $n_i=n_i^{\rm CMS}$, where $n_i^{\rm CMS}$ is the number of observed dijet events at CMS in the $i$-th dijet mass bin.~Computing $Z$ for discovery regions, we use the dataset $n_i=b_i(\boldsymbol{\theta}_{\rm bf})+s_i$, where $\boldsymbol{\theta}_{\rm bf}$ is the value of $\boldsymbol{\theta}$ that maximises $\mathscr{L}\left(\boldsymbol{0},\boldsymbol{\theta}\right)$ for $n_i=n_i^{\rm CMS}$.

\begin{table}[t]
\begin{center}
    \begin{tabular}{lccccccc}
    \toprule
 Spin 0 DM & Op. & & $g_q$ & & $g_\text{DM}$ & & $M_\text{eff}$ [GeV]\\
$\left(S \otimes S\right)_0$ &1	&&$h_1$	&&$g_1$	&& 14600 \\
$\left(S \otimes i\partial\right)_0$ &1	&&$h_3$	&&$g_4$	&& 10300	\\
\boldmidrule
 Spin 1/2 DM & Op. && $g_q$ && $g_\text{DM}$ && $M_\text{eff}$ [GeV]\\
$\left(S \otimes S\right)_{1/2}$ &1	&&$h_1$	&&$\lambda_1$	&& 14600	\\
$\left(V \otimes V\right)_{1/2}$ &1	&&$h_3$	&&$\lambda_3$	&& 7260	\\
$\left(A \otimes A\right)_{1/2}$ &4	&&$h_4$	&&$\lambda_4$	&& 147	\\
$\left(V \otimes A\right)_{1/2}$ &8	&&$h_3$	&&$\lambda_4$	&& 225	\\
$\left(S \otimes PS \right)_{1/2}$ &11	&&$h_1$	&&$\lambda_2$	&& 352 \\
\boldmidrule
Spin 1 DM & Op. && $g_q$ && $g_\text{DM}$ && $M_\text{eff}$ [GeV]\\
$\left(S \otimes S\right)_1$ &1	&&$h_1$	&&$b_1$	&& 14600	\\
$\left(V \otimes i\partial\right)_1$ &1	&&$h_3$	&&$b_5$	&& 10300	\\
\bottomrule
 \end{tabular}    	
\end{center}
\caption{Benchmark points producing 150 signal events in an idealised version of 
XENONnT for $m_\chi = 50\gev$~\cite{Baum:2017kfa}.~Consistently 
with~\cite{Baum:2017kfa}, in the case of spin 1 DM we do not consider the 
contribution to $M_{\rm eff}$ from effective operators that depend on the 
symmetric combination of polarisation vectors denoted by $\mathcal{S}$. We 
omitted models that are not compatible with a dijet signal 
in the relevant parameter space.}
\label{tab:benchmarks}
\end{table}

\section{Analysis}
\label{sec:method}
In this section we investigate the impact of a XENONnT signal on the interpretation of current dijet searches at the LHC, and on the prospects for dijet signal discovery at the HL-LHC. 

Let us start by describing our assumptions about the hypothesised XENONnT 
signal.~We assume that XENONnT with an exposure of 
$20\,\text{ton}\times\text{year}$ has detected 150 nuclear recoil events due to 
DM-nucleus scattering.~Roughly, this number of signal events corresponds to DM 
models lying just below current XENON1T limits.~We consider an idealised version 
of the XENONnT detector with infinite energy resolution, an energy threshold of 
5 keV, and 100\% detector efficiency.\footnote{ 
The detector deficiencies and resolution of XENON will have a minor impact on 
$M_{\rm eff}$, in particular compared to astrophysical uncertainties as 
discussed in more detail in 
Ref.~\cite{Baum:2017kfa}. We use the 150 signal events in the idealized version of XENON as a benchmark. A more realistic treatment of the detector would lead to less, but still $\mathcal{O}(100)$, signal events for a given benchmark point. Furthermore, including detector efficiencies and resolution would not have significant consequences for comparing the impact of a direct detection signal in dijet searches for different models.}~To obtain the number of expected signal events at XENONnT, we 
follow~\cite{dent:2015zpa} and integrate the differential rate of nuclear recoil 
events in Eq.~(\ref{eq:rate}) in the 5 to 45 keV range using 
\texttt{DMFormFactor}~\cite{anand:2013yka}. The total number of events is then 
obtained by multiplying the result by a $20\,\text{ton}\times\text{year}$ 
exposure.~Direct detection experiments are not sensitive to the individual 
parameters of the simplified models, but only to the DM mass $m_{\rm DM}$ and 
the effective mediator mass
\begin{equation}
   M_{\rm eff} \equiv \frac{m_{\rm med}}{(g_q/0.1) (g_{\rm DM}/0.1)}\;.
\label{eq:meff}
\end{equation}
Note that we chose to normalize the coupling constants to $g_q = g_{\rm DM} = 0.1$, corresponding to typical values for weak couplings, in the definition of $M_{\rm eff}$.

For the simplified models in Eq.~(\ref{eq:simpL}), Tab.~\ref{tab:benchmarks} 
shows the values of $M_{\rm eff}$ required to produce 150 signal events at an 
idealised version of XENONnT.~Notice that a signal at XENONnT would constrain 
$M_{\rm eff}$ univocally, with an associated relative uncertainty of about 
20\%~\cite{Catena:2017xqq} that would be negligible compared to astrophysical 
uncertainties.~Furthermore, experimental errors on the reconstructed value of 
$m_{\rm DM}$ are also expected to be negligible in this setup\footnote{This holds as long as the mass of the DM candidate is not much heavier than the mass of a xenon nucleus, the target in XENONnT, see e.g. Ref.~\cite{Edwards:2018lsl}.}, and we therefore 
set $m_{\rm DM}$ to its benchmark value, i.e.~$m_{\rm DM}=50$~GeV.~In addition 
to the constraints on $M_{\rm eff}$ and $m_{\rm DM}$ from the detection of 150 
signal events at XENONnT, we also require perturbative couplings $|g_{\rm DM}| < 
\sqrt{4\pi}$ and $|g_q| < \sqrt{4\pi}$.~Finally, we assume universal quark 
couplings $g_u = g_d = g_s = g_c = g_b = g_t \equiv g_q$, and negligible 
coupling of the mediator to leptons, i.e.~$g_\ell \simeq 0$, in agreement with 
current searches for resonances in dilepton final states at the 
LHC~\cite{Sirunyan:2018exx}.~Having described our assumptions about the 
hypothesised XENONnT signal, we now investigate its impact on the interpretation 
of current dijet searches at the LHC (Sec.~\ref{sec:limits}), and on the 
prospects for dijet signal discovery at the HL-LHC (Sec.~\ref{sec:discovery}).

\subsection{Impact of a XENONnT signal on LHC dijet exclusion limits}
\label{sec:limits}
For the benchmark parameters which would give rise to 150 signal events in XENONnT, we calculate 95\% C.L. exclusion limits on the mediator's coupling to quarks, $g_q$, from current searches for resonances in dijet final states at the LHC. Because we fixed the DM mass to $m_{\rm DM} = 50\,$GeV (assuming perfect mass reconstruction), our XENONnT analysis described above yields a surface of parameter points in the space spanned by $\{m_{\rm med}, g_q, g_{\rm DM}\}$, defined by the respective values of $M_{\rm eff}$ reported in Tab.~\ref{tab:benchmarks}.~On this surface, the mediator's coupling to DM, $g_{\rm DM}$, is a function of $m_{\rm med}$  and $g_q$.~Geometrically, the function $g_{\rm DM}=g_{\rm DM}(m_{\rm med}, g_q)$ can be obtained by projecting the surface defined by $M_{\rm eff}$ to the $m_{\rm med} - g_q$ plane.~In practice, for each benchmark point in Tab.~\ref{tab:benchmarks} we obtain $g_{\rm DM}$ by solving Eq.~(\ref{eq:meff}) for $g_{\rm DM}$ at each point in the $(m_{\rm med}, g_q)$ plane.

To calculate the 95\% C.L. exclusion limits on $g_q(m_{\rm med})$ arising from 
resonant dijet searches at the LHC and a signal at XENONnT, we use the profile 
likelihood ratio method outlined in Sec.~\ref{sec:stat}. For each simplified 
model, we simulate the corresponding dijet invariant mass spectrum using the 
chain of numerical programmes described in Sec.~\ref{sec:theory} on a grid in 
$(m_{\rm med}, g_q)$, setting $g_{\rm DM}$ to the value obtained from the 
$M_{\rm eff}$ constraint at each point and fixing $m_{\rm DM} = 50\,$GeV.
~After simulating the dijet mass spectrum, we integrate it to obtain 
the number of expected dijet events, $s_i$, in bins of dijet invariant mass 
labeled by the integer $i$ and of variable width, as in the high-mass search for 
narrow resonances in dijet final states performed by 
CMS~\cite{1676214}.~Following~\cite{1676214}, we assume an integrated luminosity 
of 36 fb$^{-1}$, a centre-of-mass energy of $\sqrt{s} = 13\tev$, and focus on 
the 1.6 -- 3.9 TeV range for the dijet invariant mass.
~For models where 150 events in XENONnT arise for relatively large effective 
mediator masses, e.g. $M_{\rm eff} = 1.46\tev$ for $(S \otimes S)_{1/2}$, dijet 
searches at the LHC constrain regions of parameter space where $g_q$ is sizable 
and $g_{\rm DM} \ll g_q$ (with typical values of $g_{\rm DM} \sim 
\mathcal{O}(10^{-3})$). Then, the mediator predominately decays into pairs of 
quarks. For models with much smaller $M_{\rm eff}$ corresponding to 150 signal 
events in XENONnT, e.g. $M_{\rm eff} = 147\gev$ for $(A \otimes A)_{1/2}$, 
dijet searches constrain regions of parameter space where $g_{\rm DM} \gtrsim 
g_q$ (with typical values close to the perturbativity bound, $g_{\rm DM} = 
\sqrt{4\pi}$), and hence the branching ratio of the mediator into pairs of 
quarks is suppressed.

\begin{figure*}[t]
\begin{center}
\begin{tabular}{cc}
\includegraphics[width=.5\linewidth]{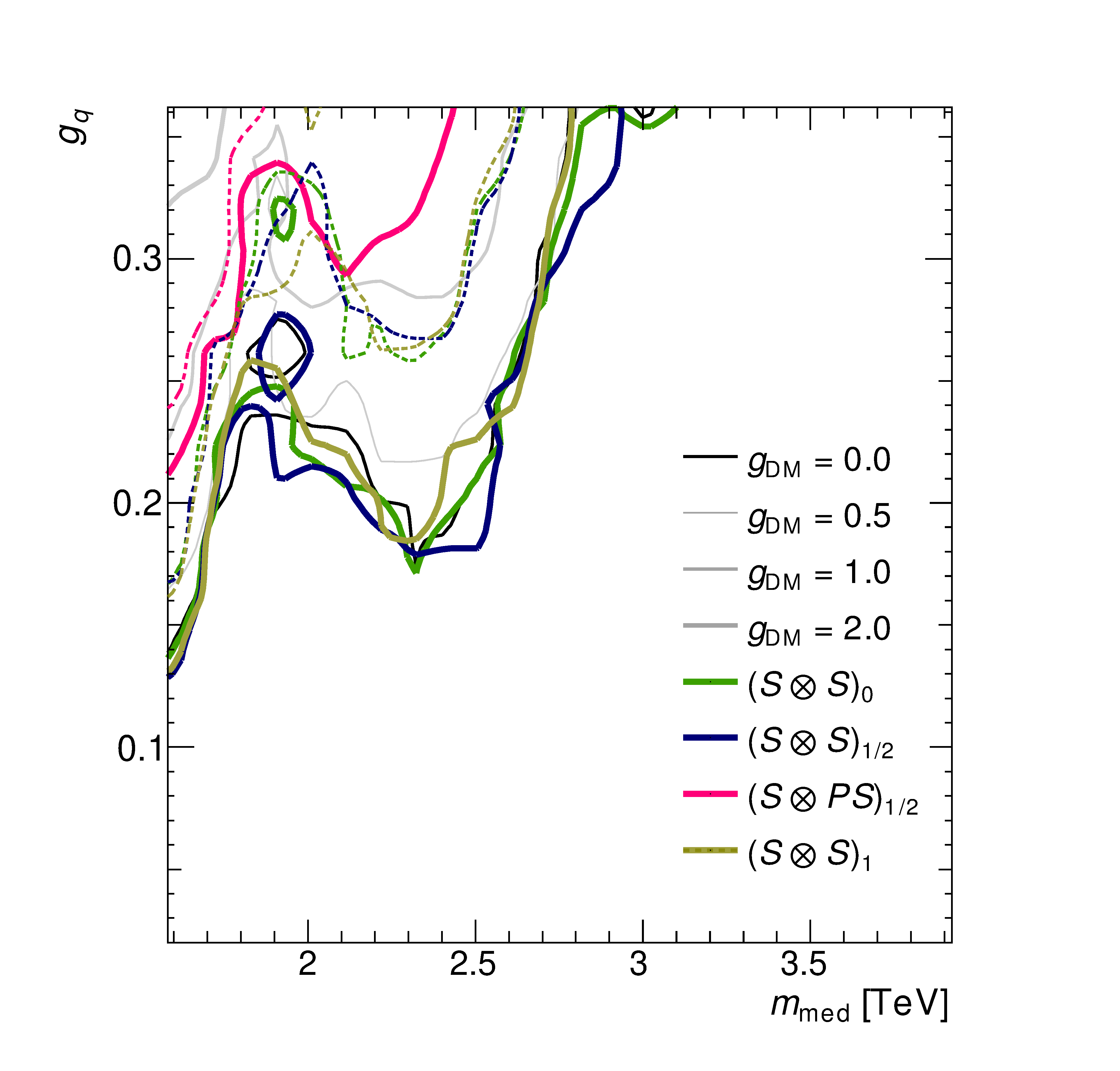}
&
\includegraphics[width=.5\linewidth]{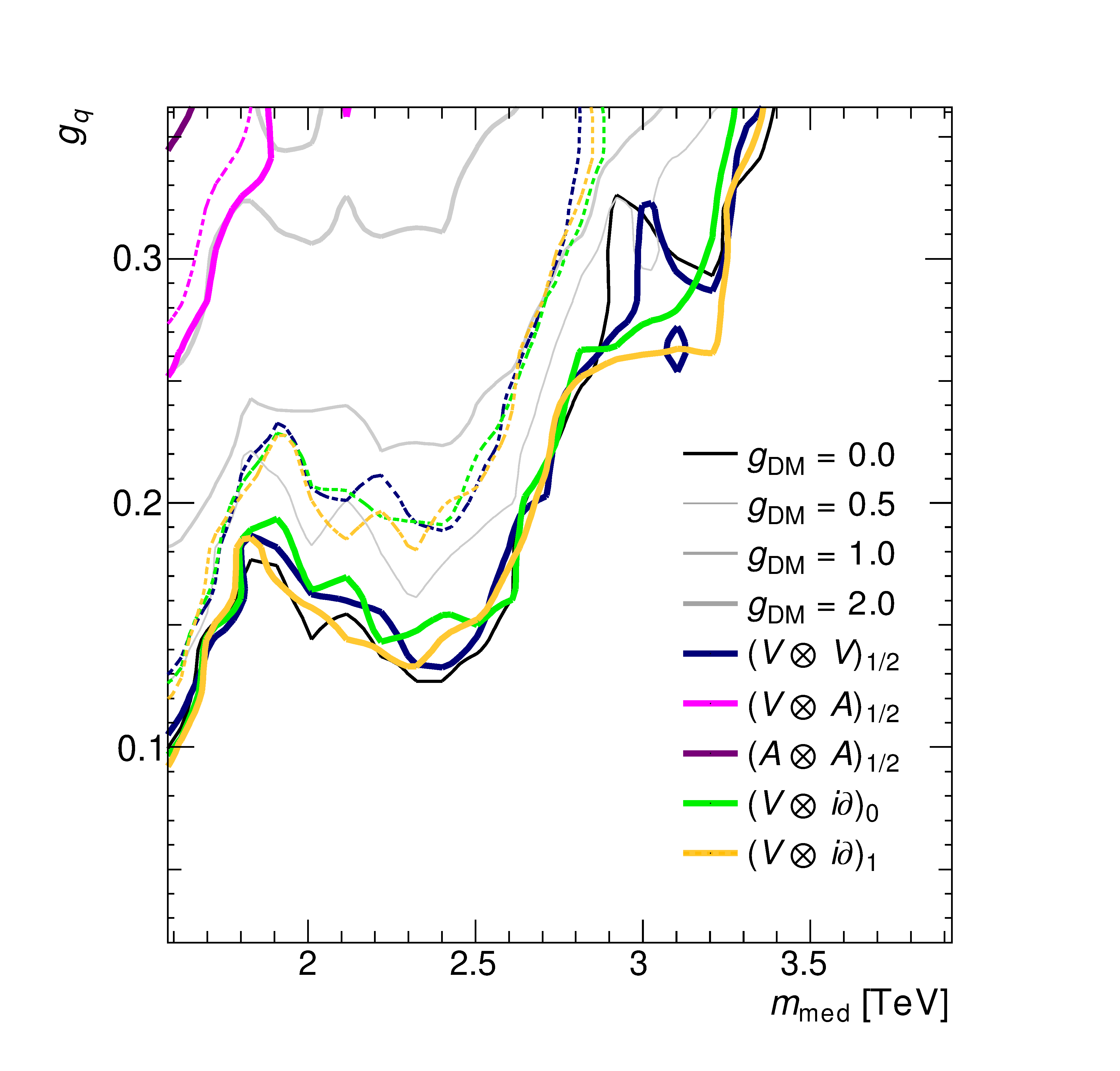}
\\
(a) & (b)
\end{tabular}
\end{center}
\caption{95\% C.L. ($3\sigma$) exclusion limits on $g_q$ from the null result 
of present searches for narrow resonances in dijet final states at the LHC 
obtained by setting $m_{\rm DM}=50$~GeV and $g_{\rm DM}$ to the value required 
by the detection of 150 signal events at XENONnT indicated by solid 
(dotted) coloured lines.~Exclusion limits are presented for simplified models 
with a scalar mediator (a) and for models with a vector mediator (b).~We show 
only the models that are not yet fully excluded by dijet searches in the 
relevant parameter range.~In both panels we used data from a CMS search for 
narrow resonances in final states involving a dijet 
corresponding to an integrated luminosity of $36 \invfb$.~For models $\left(S 
\otimes S\right)_{1/2}$ (a) and $\left(V \otimes V\right)_{1/2}$ (b), the figure 
also shows 95\% C.L. exclusion limits obtained by setting $g_{\rm DM}$ to 
XENONnT-independent values (grey lines). \label{fig:exclusion}}
\end{figure*}

Fig.~\ref{fig:exclusion} shows the impact that the detection of 150 signal 
events at XENONnT would have on the 95\% C.L. exclusion limits on $g_q$ from the 
null result of present searches for narrow resonances in dijet final states at 
the LHC.~The left panel refers to simplified models with scalar or pseudo-scalar 
mediators, whereas the right panel corresponds to simplified models with vector 
or pseudo-vector mediators.~In Fig.~\ref{fig:exclusion} we report exclusion 
limits only for a subset of the simplified models and corresponding benchmark 
points in Appendix~\ref{sec:app}.~For the models not shown, the benchmark 
points would correspond to 95\% C.L. exclusion limits extending to regions in 
parameter space where coupling constants are non-perturbative.~Specifically, 
this applies to benchmark points where $M_\text{eff} \ll m_\text{med}$.
~In Fig.~\ref{fig:exclusion} we display all the models for which combinations of parameters which would give rise to 150 signal events in XENONnT are not yet 
excluded by LHC dijet results; labeled according to 
the notation introduced in Eq.~(\ref{eq:simpL}).
~For the model $\left(S \otimes S\right)_{1/2}$ in the left panel -- fermionic 
DM and scalar mediator -- and the model $\left(V \otimes V\right)_{1/2}$ in the 
right panel -- fermionic DM and vector mediator -- we also show 95\% C.L. 
exclusion limits obtained using values for $g_{\rm DM}$ which are not related to 
constraints on $M_{\rm eff}$ from the detection of 150 signal events at XENONnT 
(grey curves).~For these two models, 150 signal events at XENONnT require 
$M_\text{eff} \gg m_\text{med}$, which implies $g_{\rm DM}\simeq 0$.~This 
explains why for models $\left(S \otimes S\right)_{1/2}$ and $\left(V \otimes 
V\right)_{1/2}$ 95\% C.L. exclusion limits computed assuming 150 signal events 
at XENONnT or setting $g_{\rm DM}=0$ are close to each other.~On the other hand, 
while exclusion limits in Fig.~\ref{fig:exclusion} depend only indirectly on 
$g_{\rm DM}$ via the total mediator decay width and branching ratio into quarks, 
a large coupling to DM can significantly reduce the branching ratio into quarks, 
and therefore lead to significantly weaker exclusion limits on $g_q$.

\subsection{Impact of a XENONnT signal on LHC dijet 5$\sigma$ discovery contours}
\label{sec:discovery}

\begin{figure*}[t]
\begin{center}
\begin{tabular}{cc}
\includegraphics[width=.5\linewidth]{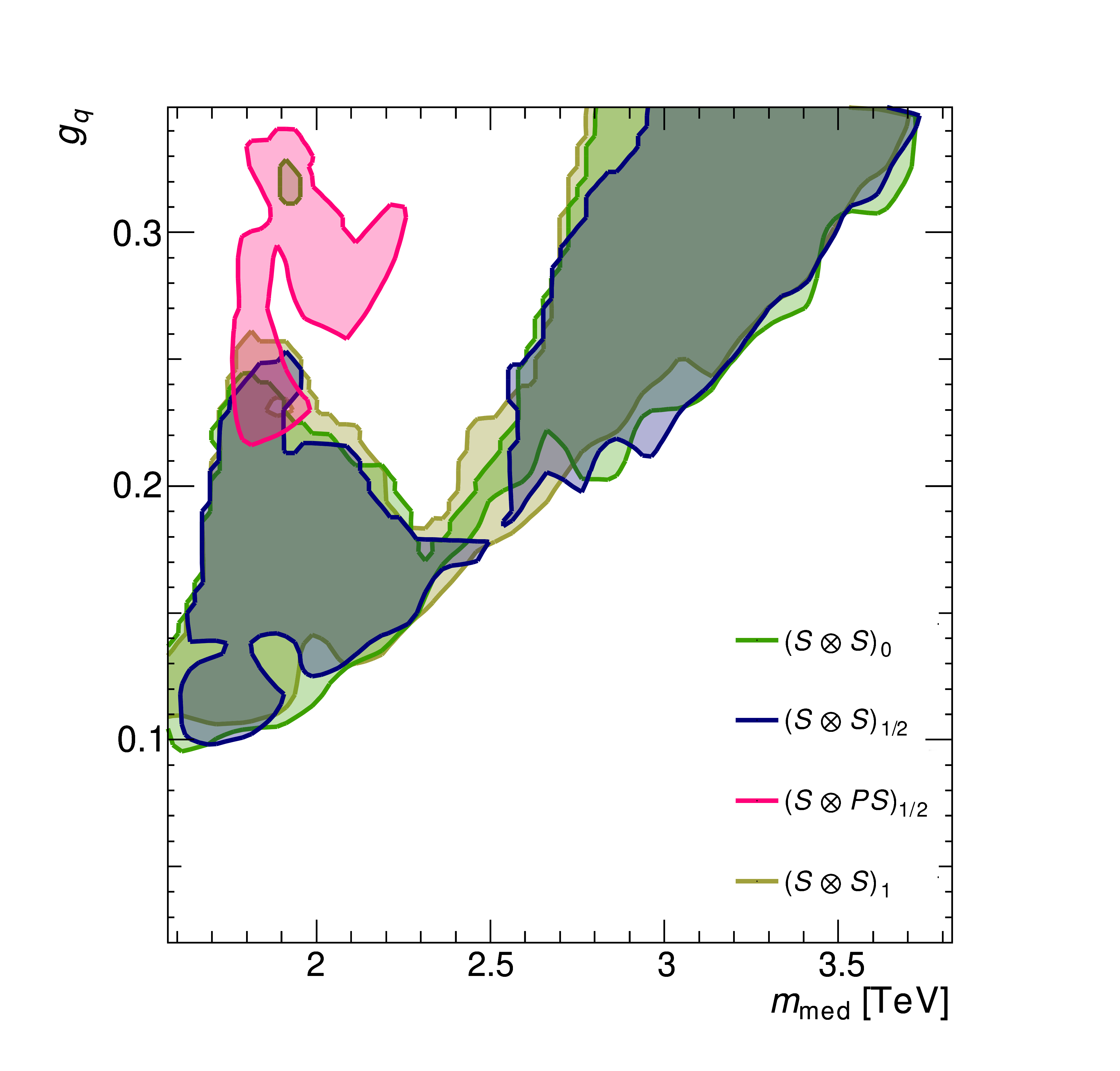}
&
\includegraphics[width=.5\linewidth]{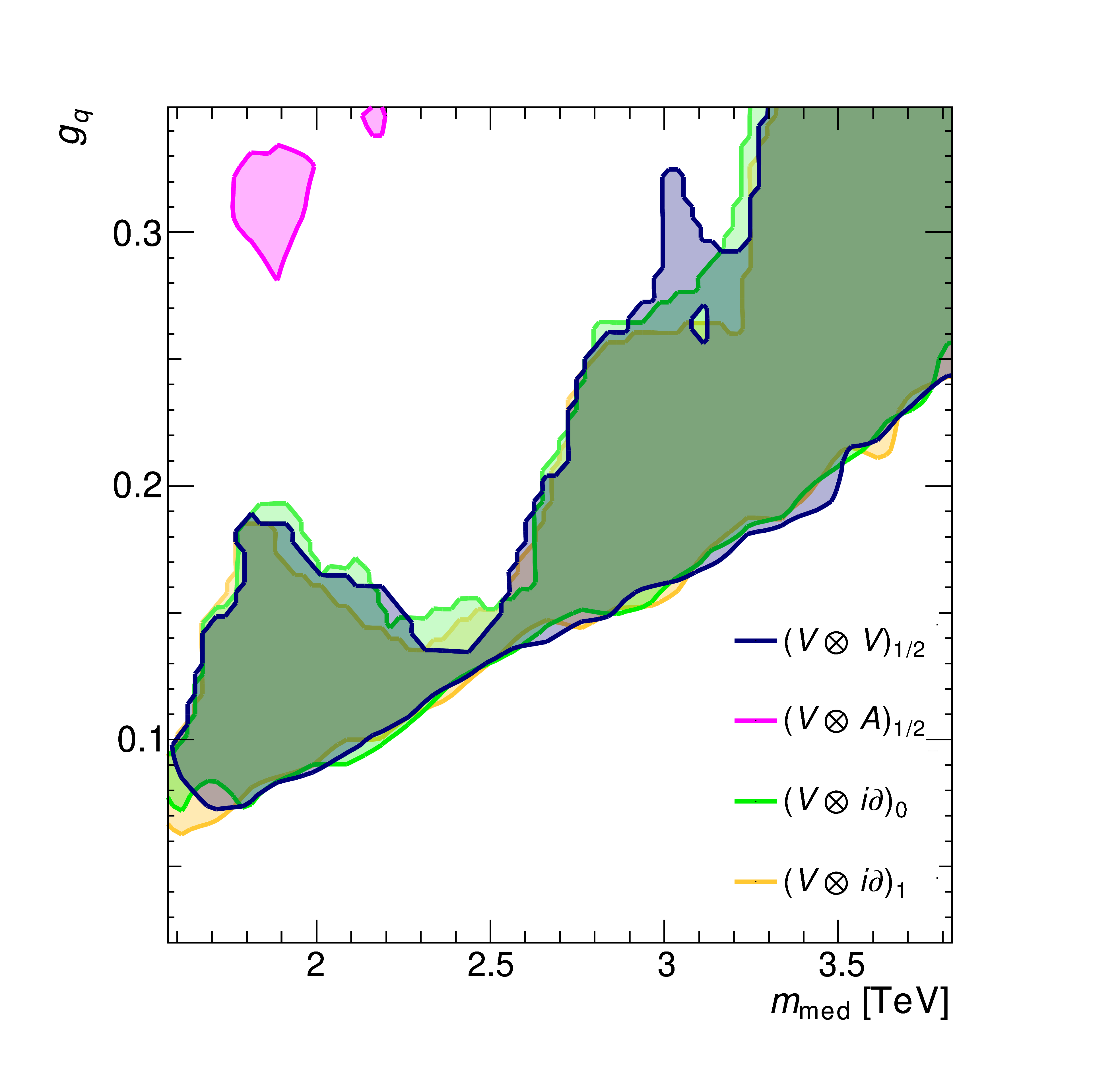}
\\
(a) & (b)
\end{tabular}
\end{center}
\caption{Regions in the $(m_{\rm med}, g_q)$ plane where a narrow 
resonance could be discovered with $Z \geq 5\sigma$ C.L. in dijet final states 
at the HL-LHC that are at the same time compatible with the detection of 
150 signal events at XENONnT. The lower boundary of the region for each model is the smallest coupling $g_q(m_{\rm med})$ for which we would expect a $5\sigma$ discovery at the HL-LHC, while the upper boundary is given by the current $95\,\%$\,C.L. exclusion limits from $36\invfb$ of CMS data.~The left panel (a) corresponds to simplified models with a 
scalar mediator, while the right panel (b) refers to models with a vector 
mediator.
\label{fig:discovery}}
\end{figure*}

\begin{figure*}[t]
\begin{center}
\begin{tabular}{cc}
\includegraphics[width=.5\linewidth]{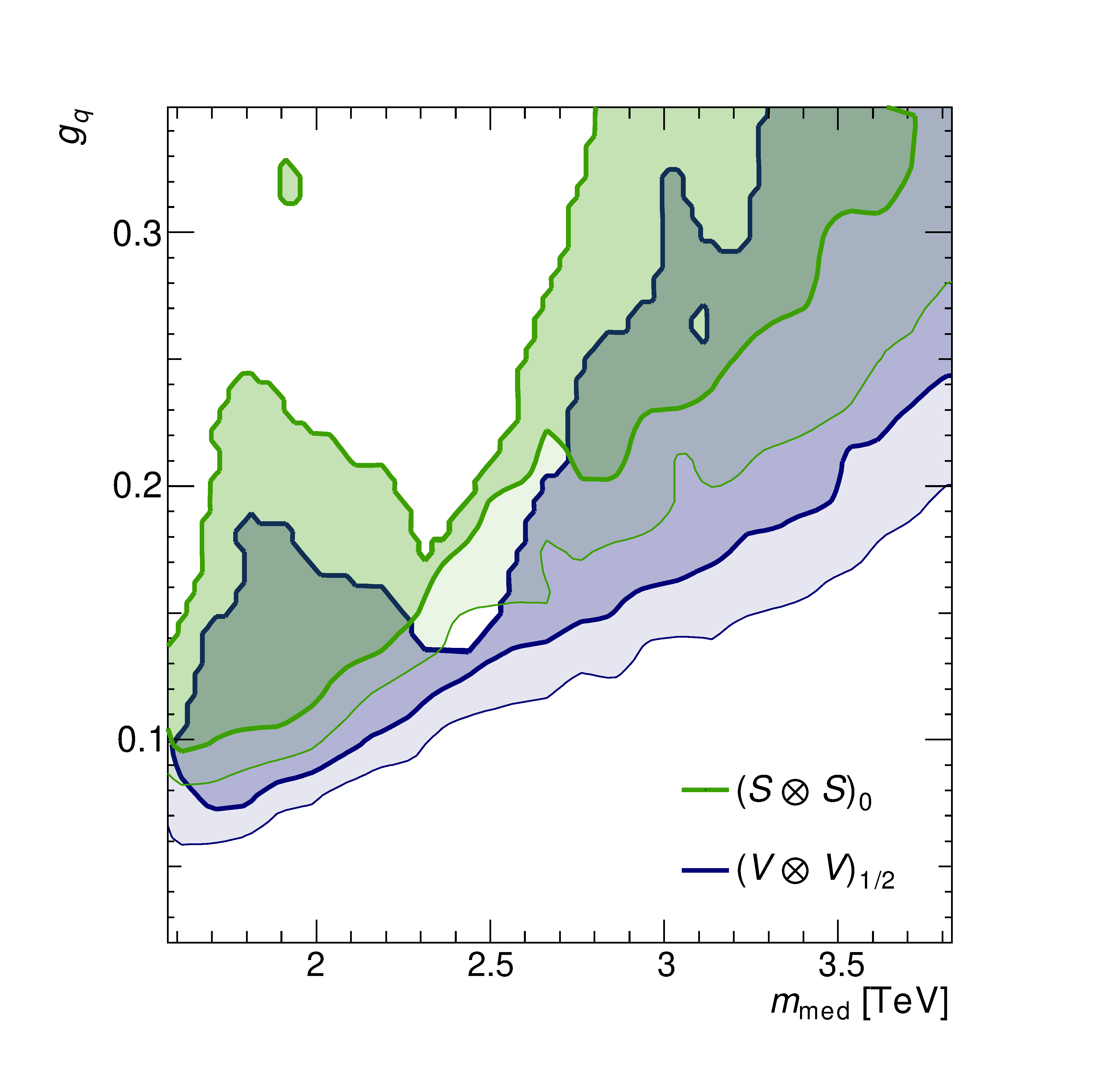}
&
\includegraphics[width=.5\linewidth]{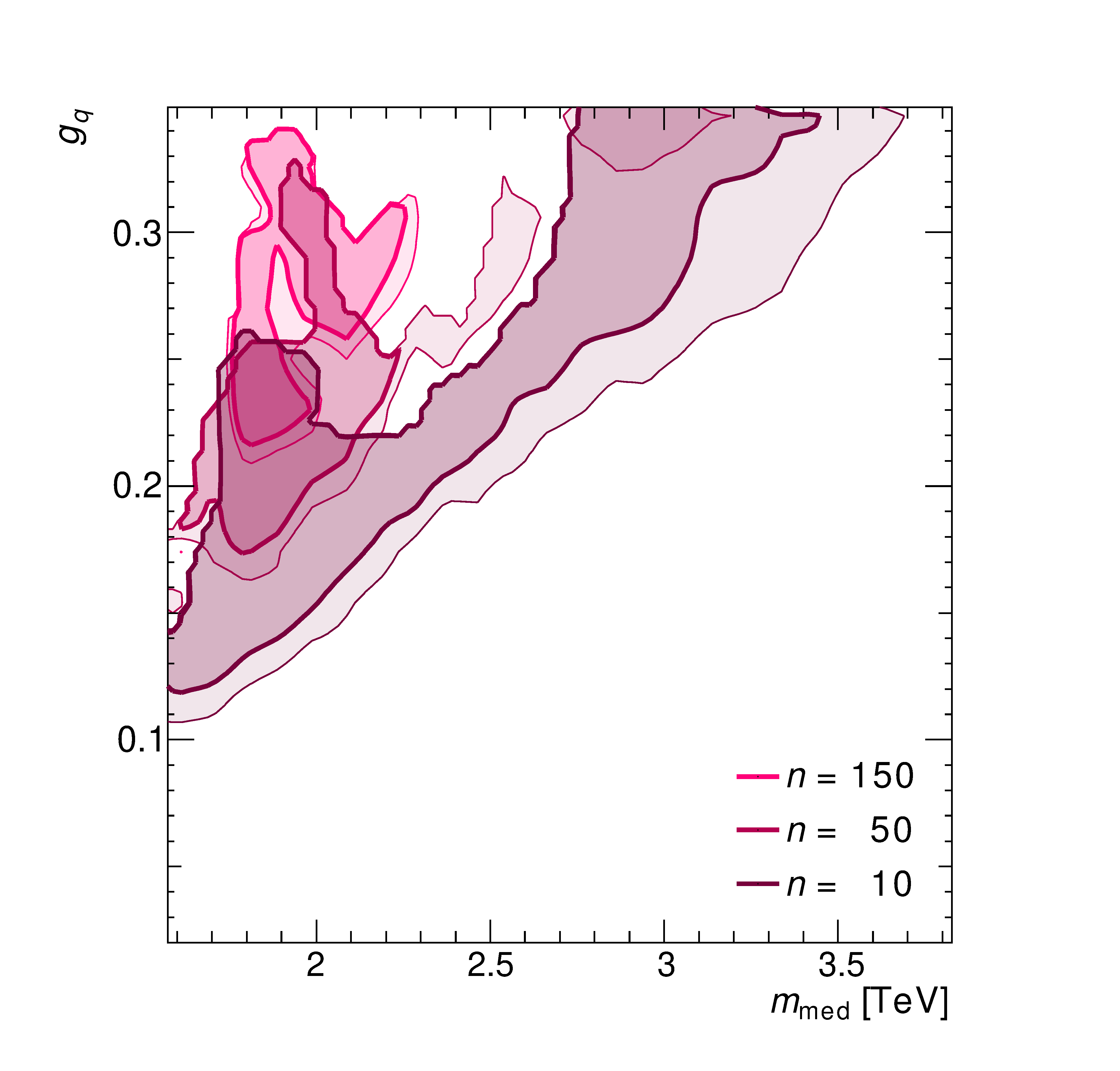}
\\
(a) & (b)
\end{tabular}
\end{center}
\caption{The left panel (a) is the same as Fig.~\ref{fig:discovery}, but 
now focusing on the models $\left(S \otimes S\right)_0$ and $\left(V \otimes 
V\right)_{1/2}$.~Both models generate $\hat{\mathcal{O}}_1$ as a leading 
non-relativistic operator for DM-nucleon interactions.~
The thick (thin) outline corresponds to a $5\sigma$ ($3\sigma$) 
discovery.~Discovery contours are partly non overlapping.
In the right panel (b) we show how the regions where dijets searches could make 
a discovery in the $\left(S \otimes PS\right)_{1/2}$ model if 
instead of $n=150$ only $n=50$ or $n=10$ events would be observed at XENONnT.
\label{fig:compare}}
\end{figure*}

In this section, we investigate the impact of a XENONnT signal on the prospects 
for dijet signal discovery at the HL-LHC.~We use the profile likelihood ratio 
method outlined in Sec.~\ref{sec:stat} to identify the contours in the $(m_{\rm 
med}, g_q)$ plane where simultaneously:~1) a narrow resonance in dijet final 
states at the HL-LHC could be discovered with a statistical significance of 
$5\sigma$;~2)150 signal events are expected at XENONnT.~As before, 
we set $m_{\rm DM}$ to 50~GeV and extract $g_{\rm DM}$ from $M_{\rm eff}$ using 
the XENONnT input.~Concerning dijet signal and background calculation, as well 
as our choice of likelihood function and dataset, we proceed as described in 
Sec.~\ref{sec:stat}.

Fig.~\ref{fig:discovery} shows the regions in the $(m_{\rm med}, g_q)$ plane 
where a dijet signal could be discovered at the HL-LHC with a statistical 
significance larger than or equal to $5\sigma$, and which are at the same time 
compatible with the detection of 150 signal events at XENONnT.~Regions with 
different colours correspond to distinct simplified models in 
 Eq.~(\ref{eq:simpL}). In each region, the lower boundary $g_q^{\rm 
min}(m_{\rm med})$ is the smallest coupling $g_q$ for which the corresponding 
model could be discovered with a significance of $5\sigma€$ in 
dijet searches at the HL-LHC. The upper 
boundary $g_q^{\rm max}(m_{\rm med})$ is given by the 95\% C.L, exclusion limits 
from $36\invfb$ of data discussed in Sec~\ref{sec:limits}, 
cf. Fig.~\ref{fig:exclusion}.~The left panel corresponds to simplified models 
with a scalar or pseudo-scalar mediator, while the right panel refers to models 
with vector or pseudo-vector mediators.~We use the labels introduced in the 
previous sections.~Models that do not appear in 
Fig.~\ref{fig:discovery} are not compatible with the simultaneous discovery of 
150 signal events at XENONnT and the $5\sigma$ detection of a dijet signal at 
the HL-LHC. Note that in case of no signal discovery the $5\sigma$ lines in 
Fig.~\ref{fig:discovery} would roughly correspond to the corresponding HL-LHC 
exclusion limits with the same significance.

Interestingly, we find that only a subset of models would actually be compatible 
with the simultaneous detection of a signal at XENONnT and at the 
HL-LHC.~Furthermore, we find that some of the models in Fig.~\ref{fig:discovery} 
can potentially be distinguished in dijet searches, since the mediator mass can 
approximately be reconstructed from an analysis of the dijet invariant mass at 
the HL-LHC.~For example, the discovery of a dijet signal for a mediator mass 
$m_\text{med}\gtrsim 2.5 \tev$ would exclude the models $\left(S \otimes PS\right)_{1/2}$ and 
$\left(V \otimes A\right)_{1/2}$, leaving only models with $\hat{\mathcal{O}}_1$ as leading 
non-relativistic operator for DM-nucleon interactions.~On the other hand, the 
discovery of a dijet signal at lower mediator masses would also be compatible 
with models $\left(S \otimes PS\right)_{1/2}$ and $\left(V \otimes A\right)_{1/2}$, which in the 
non-relativistic limit generate the interaction operators 
$\hat{\mathcal{O}}_{11}$ and $\hat{\mathcal{O}}_8$, respectively. Note, that the 
regions shown in Fig.~\ref{fig:discovery} do not take into account limits on 
simplified models which can be obtained from the preliminary results published 
by the CMS collaborations from $78\invfb$ of data~\cite{CMS-PAS-EXO-17-026} presented 
in~\cite{Baum:2018sxd}. These results rule out all of the regions for the 
models $\left(S \otimes PS\right)_{1/2}$ and $\left(V \otimes A\right)_{1/2}$ shown in 
Fig.~\ref{fig:discovery} at 95\% C.L. However, a $5\sigma$ discovery could still 
be made at the HL-LHC for such models if one loosens the assumption of 150 
signal events being produced at XENONnT. Regarding the models with couplings
$[h_3,\Re{(b_7)}]$ and $[h_3,\Im{(b_7)}]$, a spin 1 mediator with a spin 1 DM 
candidate, cf. Eq.~\eqref{eq:L11}, we did not compute them 
explicitly. The benchmark value for the effective mediator mass giving rise to 
150 events in XENONnT is $M_{\rm eff} \sim 200\gev$, thus, one would expect 
similar regions as in the $\left(V \otimes A\right)_{1/2}$ scenario. However, 
the partial width for a spin 1 mediator decaying via $b_7$ is enhanced by a factor 
$\sim(m_\text{med}/m_\text{DM})^2$ with respect to a decay via 
$\lambda_3$ in the $\left(V \otimes A\right)_{1/2}$ model~\cite{Catena:2017xqq}. Thus, the branching ratio into quarks is 
suppressed and we do not expect a significant chance to discover such a scenario 
at the HL-LHC.

As an aside comment, we mention here that $g_q$ could in principle be inferred from the measurement of the mediator decay width, assuming that $m_\text{med}$ and $M_\text{eff}$ are both known.~However, this would require a very accurate measurement of the dijet invariant mass spectrum to extract the mediator decay width from data collected at the HL-LHC.~Most likely, the number of signal events recorded in the initial stages of the HL-LHC would not suffice to reconstruct the invariant mass spectrum with the precision required to indirectly infer $g_q$.

In the left panel of Fig.~\ref{fig:compare} we compare the two models $\left(S 
\otimes S\right)_0$ and $\left(V \otimes V\right)_{1/2}$ in more detail.~We find 
that these models predict partly non overlapping contours in the $(m_{\rm med}, 
g_q)$ plane.~This is an interesting result, since it shows that models 
generating $\hat{\mathcal{O}}_1$ as leading non-relativistic operator can in 
principle be discriminated if a dijet signal is observed at the 
HL-LHC.\footnote{
The island like structures in Fig.~\ref{fig:compare} as well as the 
 $3\sigma$\,C.L exclusion limit shown in Fig.~\ref{fig:exclusion} 
(dotted lines) indicate that the uncertainties in the upper boundary of the 
region (corresponding to the current $95\,\%$\,C.L exclusion limit) are larger 
than the separation of the models. If we focus on the lower boundary, 
however, we see that there is a remaining region where a signal could only be 
discovered with $5\sigma$ significance for the model $\left(V \otimes 
V\right)_{1/2}$ and not for $\left(S \otimes S\right)_0$ thus allowing in 
principle for a separation of the models in part of the parameter space.
}
~This is in contrast with what we found investigating the impact of a XENONnT 
signal on LHC monojet searches~\cite{Baum:2017kfa}.~In that work, we found that 
models generating $\hat{\mathcal{O}}_1$ as leading operator in the 
non-relativistic limit are not observable in monojet searches at the LHC if the 
model parameters are such that $\mathcal{O}(100)$ signal events would be 
observed at 
XENONnT~\cite{Baum:2017kfa}.

Let us now investigate the dependence of our results on the number of signal events observed at XENONnT, $n$.~\reffig{fig:compare}, right panel, shows how discovery regions change if instead of $n=150$ signal events, only $n=50$ or $n=10$ events are observed at XENONnT. As an example, we show results for the model $\left(S \otimes PS\right)_{1/2}$. Note that the effective mass reconstructed from XENONnT scales with the number of events as $M_\text{eff} \sim n^{-4}$. For a fixed combination of parameters $(m_\text{med},g_q)$ this implies the scaling 
$g_\text{DM}\sim\sqrt{n}$. 
A larger number of events at XENONnT therefore implies a larger partial decay 
width of the mediator into DM for a given $g_q$ and mediator mass, which 
weakens the significance of a potential dijet signal. Fewer events at XENONnT 
instead lead to the predominant decay into SM particles leading to a stronger 
dijet signal in comparison, which is limited by the ideal case where 
$g_{\rm DM}\cong0$ and the mediator decays exclusively into SM particles.

As a consequence, the exclusion limits as well as the $5\sigma$ discovery 
contour move towards higher masses and smaller quark couplings. Furthermore, the 
overall region where a discovery at the HL-LHC is possible becomes larger. We 
expect a similar behaviour for the model $(V \otimes A)_{1/2}$. For the remaining 
models shown in Fig.~\ref{fig:discovery}, the mediator's decay width is 
dominated by the partial width corresponding to decays into quarks. Therefore, a 
smaller number of events observed at XENONnT implying a smaller coupling $g_{\rm 
DM}$ for fixed $g_q$ and $m_{\rm med}$ would have virtually no impact on the 
mediator's decay width and branching ratios. Thus, the regions in which such 
models could give rise to a $5\sigma$ discovery at the HL-LHC are nearly 
independent of the number of events observed at XENONnT.

Let us also qualitatively investigate what impact a DM mass different from 
$m_\text{DM} = 50\,$GeV, would have on our results. The detector can resolve 
dark matter masses up to about $100\,$GeV reasonably well, see, e.g., 
Refs.~\cite{Catena:2017xqq,Edwards:2018lsl}. For larger masses the direct 
detection signal becomes largely independent of $m_\text{DM}$.  Due to this 
degeneracy, XENONnT is not expected to achieve an accurate reconstruction of 
dark matter masses above $\sim 100\,$GeV. We therefore compare the mass 
$m_{\rm DM} = 50\,$GeV, where XENONnT is most sensitive, with $m_{\rm DM} = 
100\,$GeV.
We restrict our discussion to the models $\left(S \otimes PS\right)_{1/2}$ and 
$\left(V \otimes A\right)_{1/2}$ for which prospects at the HL-LHC depend 
strongly on the number of events observed at XENONnT, $n$. 
Naively, one would assume that the dijet production cross section depends on 
the DM mass via the branching ratios of the mediator. However, the dependence on 
the mediator mass, $m_{\rm med}$, is much stronger, yielding virtually 
unchanged discovery regions when assuming $m_{\rm DM} = 100\,$GeV 
instead of $m_{\rm med} = 50\,$GeV. On the other hand, DM direct detection 
experiments using xenon as target material can probe the smallest WIMP-nucleon 
scattering cross sections for DM masses of $m_{\rm} \sim 50\,$GeV. Thus, 
observing 150 events at XENONnT corresponds to larger WIMP-nucleus cross 
sections, and in turn smaller $M_{\rm eff}$, for larger $m_{\rm DM}$. Since 
smaller $M_{\rm eff}$ correspond to larger $g_{\rm DM}$ for fixed values of 
$m_{\rm med}$ and $g_q$, increasing $m_{\rm DM}$ to values larger than $50\,$GeV 
has the opposite effect as a smaller number of events observed at XENONnT 
discussed above. Similar to that case, the discovery regions presented for the 
other models considered here are expected to remain approximately unchanged when 
for example assuming $m_{\rm DM} = 100\,$GeV instead of $m_{\rm med} = 50\,$GeV.

Finally, we would like to stress that the two models $\left(S \otimes PS\right)_{1/2}$ and $\left(V \otimes A\right)_{1/2}$ can simultaneously be observed in monojet~\cite{Baum:2017kfa} and dijet searches at the LHC (see Fig.~\ref{fig:discovery}).~For model $\left(S \otimes PS\right)_{1/2}$, a dijet signal is only observable close to $m_\text{med} \approx  2 \tev$, $g_q \approx 0.3$ and $g_\text{DM} \approx 1$, as it can be inferred from Fig.~\ref{fig:discovery}.~Interestingly, these models also generate non-relativistic operators for DM-nucleon interactions, $\hat{\mathcal{O}}_{11}$ and $\hat{\mathcal{O}}_{8}$, respectively, which can statistically be discriminated from an analysis of the associated nuclear recoil  energy spectra~\cite{Baum:2017kfa}.~These models can therefore be very effectively constrained from a combined analysis of LHC and XENONnT data. 

\section{Conclusion}
\label{sec:conclusion}
In this work we have investigated the impact that a signal at XENONnT would have on the interpretation of current dijet searches at the LHC, and on the prospects for dijet signal discovery at the High-Luminosity LHC in the framework of simplified models.~In the analysis, we have focused on simplified models where DM can have spin 0, 1/2 or 1, and primarily interacts with quarks through the exchange of scalar, pseudo-scalar, vector, or pseudo-vector mediators.

Assessing the impact of a XENONnT signal on the interpretation of current dijet searches at the LHC, we have calculated 95\% C.L. exclusion limits on the coupling constant associated with the mediator-quark-quark vertex, $g_q$, as a function of the mediator mass, $m_{\rm med}$, from the null result of current searches for resonances in dijet final states at the LHC.~We have performed this calculation for the simplified models described above (and in greater detail in Appendix~\ref{sec:app}), setting $m_{\rm DM}$ to the benchmark value of $50$~GeV, and taking into account the constraint on the effective mediator mass $M_{\rm eff}$ [defined in Eq.~\eqref{eq:meff}] arising from the detection of 150 signal events at XENONnT.~The 95\% C.L. exclusion limits on $g_q$ presented here have been calculated using the standard profile likelihood method~\cite{cowan:2010js}.~We have found that for models for which a XENONnT signal implies $M_\text{eff} \ll m_\text{med}$ (in the range of $m_\text{med}$ values that we have considered), 95\% C.L. exclusion limits extend to regions in parameter space where coupling constants are non-perturbative, and therefore become trivial, because of DM detection at XENONnT.~At the same time, we have found that models for which 150 signal events at XENONnT require $M_\text{eff} \gg m_\text{med}$ are characterised by the constraint $g_{\rm DM}\simeq 0$.~In general, we have found that while exclusion limits in the $(m_{\rm med}, g_q)$ plane depend only indirectly on $g_{\rm DM}$ via the total mediator decay width and branching ratio into quarks, a large coupling to DM [i.e.~$g_{\rm DM} \sim \mathcal{O}(1)$] can significantly reduce the branching ratio into quarks, and therefore lead to significantly weaker exclusion limits on $g_q$.

Assessing the impact of a XENONnT signal on the prospects for dijet signal discovery at the HL-LHC, we have identified the contours in the $(m_{\rm med}, g_q)$ plane where a narrow resonance could be discovered with a statistical significance of $5\sigma$ in dijet final states at the HL-LHC, and which are at the same time compatible with the detection of 150 signal events at XENONnT.~Interestingly, we have found that only a subset of the simplified models in Appendix~\ref{sec:app} would actually be compatible with the simultaneous detection of a signal at XENONnT and at the HL-LHC.~We have also found that some of the models for which the two signals are compatible can potentially be distinguished if the mediator mass is approximately reconstructed from an analysis of the dijet invariant mass at the HL-LHC.~Finally, we have found that models generating $\hat{\mathcal{O}}_1$ as the leading non-relativistic operator (i.e.~canonical spin independent interactions) can in principle be discriminated if a dijet signal is observed at the HL-LHC.~Notably, in a previous work~\cite{Baum:2017kfa} we have found that the same models cannot be discriminated by combining a signal at XENONnT with the LHC monojet searches.

Ultimately, our work has explored a new aspect of the well-known complementarity between DM searches at direct detection experiments and at the LHC.~The results obtained in this study will be especially useful if DM will be discovered at XENONnT, but the methods illustrated here can in principle be applied to other combinations of  DM search experiments.

\section*{Acknowledgments}
We would like to thank Jan Conrad and Katherine Freese for their contribution to the early stages of this project.~We are grateful to Caterina Doglioni and Felix Kahlhoefer for useful insights into how to model dijet signal and background events at the LHC.~Finally, it is a pleasure to thank Kåre Fridell and Vanessa Zema for helpful discussions on non-relativistic effective theories and simplified models for dark matter.~This work is partly performed within the Swedish Consortium for Dark Matter Direct Detection (SweDCube), and was supported by the Knut and Alice Wallenberg Foundation (PI, Jan Conrad) and by the Vetenskapsr\aa det (Swedish Research Council) through contract No. 638-2013-8993 (PI, Katherine Freese).

\appendix
\section{Lagrangians for simplified dark matter models}
\label{sec:app}

In this appendix we list the Lagrangians that we considered in the analyses of Sec.~\ref{sec:limits} and Sec.~\ref{sec:discovery}~\cite{dent:2015zpa,Baum:2017kfa}.~Each Lagrangian listed here describe more than one simplified model.~By construction, simplified models are characterised by DM particle mass, $m_{\rm DM}$, mediator mass $m_{\rm med}$, and just two coupling constants:~one for a quark-quark-mediator vertex, $g_q$, and one for a DM-DM-mediator vertex, $g_{\rm DM}$.~There are no other interaction vertices in a simplified model.~For example, the simplified model associated with fermionic DM of mass $m_\chi$ and vector mediator of mass $m_G$, has $g_q=h_3$ and $g_{\rm DM}=\lambda_3$ as only coupling constants different from zero.~In all numerical applications, we assumed a universal quark-mediator coupling.

\subsection{Scalar dark matter $S$}
\paragraph{Scalar and pseudoscalar mediator $\phi$:}
\begin{eqnarray}
\mathcal{L}_{S\phi q} &=& \partial_\mu S^\dagger\partial^\mu S - m_S^2S^\dagger 
S - \frac{\lambda_S}{2}(S^\dagger S)^2 \nonumber\\
&+&\frac{1}{2}\partial_\mu\phi\partial^\mu\phi - \frac{1}{2}m_\phi^2\phi^2 
-\frac{m_\phi\mu_1}{3}\phi^3-\frac{\mu_2}{4}\phi^4 \nonumber\\
&+& \ii\bar{q}\slashed{D} q - m_q \bar q q \nonumber\\
&-&g_1m_SS^\dagger S\phi -\frac{g_2}{2}S^\dagger S\phi^2-h_1\bar q 
q\phi-ih_2\bar{q}\gamma^5q\phi\,. \label{eq:Sphi}
\end{eqnarray}
Here, $m_S$ plays the role of $m_{\rm DM}$ and $m_\phi$ that of $m_{\rm med}$. 
$\lambda_S$ is a dimensionless self-coupling of $S$ and the $\mu_i$ are 
dimensionless self-couplings of the mediator. For the purposes of this work, we 
set $\lambda_S = \mu_1 = \mu_2 = 0$. The $g_i$ are dimensionless couplings 
between $\phi$ and $S$. The couplings $h_i$ between the mediators and quarks and 
the quark mass matrix $m_q$ should in general be understood as $(6\times6)$ 
matrices and the quarks fields as vectors $q = (u,d,c,s,t,b)$ in flavour space. 
Throughout this work we assume universal (diagonal) couplings of the mediators 
to quarks such that the $h_i$ can be treated as a single number. The quark mass 
matrix $m_q$ can be assumed to be diagonal.

\paragraph{Vector and axial-vector mediator $G_\mu$:}
\begin{eqnarray}
\mathcal{L}_{SGq} &=& \partial_{\mu}S^{\dagger}\partial^{\mu}S -m_S^2 
S^{\dagger}S -\frac{\lambda_S}{2}(S^{\dagger}{S})^2  \nonumber \\
&-&\frac{1}{4}\mathcal{G}_{\mu\nu}\mathcal{G}^{\mu\nu} + 
\frac{1}{2}m_G^2G_{\mu}G^{\mu} -\frac{\lambda_G}{4}(G_{\mu}G^{\mu})^2 \nonumber 
\\
&+&i\bar{q}\slashed{D}q -m_q\bar{q}q  \nonumber\\
&-&\frac{g_3}{2}S^{\dagger}SG_{\mu}G^{\mu} 
-ig_4(S^{\dagger}\partial_{\mu}S-\partial_{\mu}S^{\dagger}S)G^{\mu} \nonumber\\
&-&h_3(\bar{q}\gamma_{\mu}q	)G^{\mu}-h_4(\bar{q}\gamma_{\mu}\gamma^5q)G^{\mu}\,. \label{eq:SG}
\end{eqnarray}
Here, ${\cal G}_{\mu\nu}$ is the field strength tensor of $G_\mu$, $m_G$ plays the role of $m_{\rm med}$, and $\lambda_G$ is a dimensionless self-coupling of $G_\mu$ which we set to zero for the purposes of this work. The $g_i$ are the dimensionless couplings of $S$ to $G_\mu$, and the $h_i$ are the couplings of $G_\mu$ to quarks. As before, the $h_i$ are in general $(6\times6)$ matrices in flavor space, but can be treated as single numbers for the universal quark coupling assumed here. The remaining parameter are as in Eq.~\eqref{eq:Sphi}.

\subsection{Fermionic dark matter $\chi$}
\paragraph{Scalar and pseudoscalar mediator $\phi$:}
\begin{eqnarray}
\mathcal{L}_{\chi\phi q} &=& \ii\bar{\chi}\slashed{D}\chi - 
m_{\chi}\bar{\chi}\chi \nonumber\\
&+&\frac{1}{2}\partial_\mu\phi\partial^\mu\phi - \frac{1}{2}m_\phi^2\phi^2 
-\frac{m_\phi\mu_1}{3}\phi^3-\frac{\mu_2}{4}\phi^4 \nonumber\\
&+& \ii\bar{q}\slashed{D} q - m_q \bar q q \nonumber\\	
&-&\lambda_1\phi\bar{\chi}\chi 
-i\lambda_2\phi\bar{\chi}\gamma^{5}\chi-h_1\phi\bar q 
q-ih_2\phi\bar{q}\gamma^5q\,. \label{eq:phichi}
\end{eqnarray}
Here, $m_\chi$ plays the role of $m_{\rm DM}$. The $\lambda_i$ are the dimensionless couplings between $\phi$ and $\chi$. The remaining parameter are as in Eq.~\eqref{eq:Sphi}.

\paragraph{Vector and axial-vector mediator $G_\mu$:}
\begin{eqnarray}
\mathcal{L}_{\chi Gq} &=& \ii\bar{\chi}\slashed{D}\chi - m_\chi\bar{\chi}\chi 
\nonumber\\
&-&\frac{1}{4}\mathcal{G}_{\mu\nu}\mathcal{G}^{\mu\nu}+\frac{1}{2}m_{G}^2G_{\mu}
G^{\mu}\nonumber\\
&+& \ii\bar{q}\slashed{D} q - m_q \bar q \nonumber\\
&-&\lambda_{3}\bar\chi\gamma^\mu\chi 
G_{\mu}-\lambda_{4}\bar\chi\gamma^\mu\gamma^5\chi G_{\mu}\nonumber\\
&-&h_3\bar{q}\gamma_{\mu}qG^{\mu}-h_4\bar{q}\gamma_{\mu}\gamma^{5}qG^{\mu}\,.
\end{eqnarray}
Beyond the parameters appearing in Eqs.~\eqref{eq:SG} and~\eqref{eq:phichi}, the $\lambda_i$ are the dimensionless couplings between $\chi$ and $G_\mu$.

\subsection{Vector dark matter $X_\mu$}
\paragraph{Scalar and pseudoscalar mediator $\phi$:}
\begin{eqnarray}
\mathcal{L}_{X\phi 
q}&=&-\frac{1}{2}{\mathcal{X}}_{\mu\nu}^{\dagger}\mathcal{X}^{\mu\nu}+m_{X}^2X_{
\mu}^{\dagger}X^{\mu}-\frac{\lambda_{X}}{2}(X_{\mu}^{\dagger}X^{\mu})^2 
\nonumber \\
&+&\frac{1}{2}(\partial_{\mu}\phi)^2-\frac{1}{2}m_{\phi}^2\phi^2-\frac{m_\phi 
\mu_1}{3}\phi^3-\frac{\mu_2}{4}\phi^4 \nonumber\\
&+&\ii\bar{q}\slashed{D}q-m_{q}\bar{q}q \nonumber\\
&-&b_1m_X\phi 
X_{\mu}^{\dagger}X^{\mu}-\frac{b_{2}}{2}\phi^2X_{\mu}^{\dagger}X^{\mu}  
\nonumber\\ 
&-&h_1\phi\bar{q}q-ih_2\phi\bar{q}\gamma^{5}q\,. \label{eq:Xphi}
\end{eqnarray}
Here, ${\cal X}_{\mu\nu}$ is the field strength tensor of $X_\mu$, $m_X$ plays the role of $m_{\rm DM}$, and $\lambda_X$ a dimensionless self-coupling of $X_\mu$ which we set to zero for the purposes of this work. The $b_i$ are dimensionless couplings of $\phi$ to $X_\mu$. The remaining parameters are as in Eq.~\eqref{eq:Sphi}.

\paragraph{Vector and axial-vector mediator $G_\mu$:}
\begin{eqnarray}
\mathcal{L}_{XGq}&=& 
-\frac{1}{2}\mathcal{X}^{\dagger}_{\mu\nu}\mathcal{X}^{\mu\nu}+m_{X}^2X^{\dagger
}_{\mu}X^{\mu}-\frac{\lambda_{X}}{2}(X_{\mu}^{\dagger}X^{\mu})^2 \nonumber\\
&-&\frac{1}{4}\mathcal{G}_{\mu\nu}\mathcal{G}^{\mu\nu}+\frac{1}{2}m_{G}
^2G_\mu^2-\frac{\lambda_G}{4}(G_\mu G^\mu)^2\nonumber \\
&+&i\bar{q}\slashed{D}q-m_{q}\bar{q}q\nonumber\\
&-&\frac{b_3}{2}G_\mu^2(X^{\dagger}_\nu X^{\nu}) 
-\frac{b_{4}}{2}(G^{\mu}G^{\nu})(X^{\dagger}_{\mu}X_{\nu}) \nonumber\\ 
&-&\left[ib_{5}X_{\nu}^{\dagger}\partial_{\mu}X^{\nu}G^\mu+b_{6}X_{\mu}^{\dagger
}\partial^\mu X_{\nu}G^{\nu}\right.\nonumber\\ 
&+& 
\left.b_{7}\epsilon_{\mu\nu\rho\sigma}(X^{\dagger\mu}\partial^{\nu}X^{\rho})G^{
\sigma} +\hc \right]\nonumber\\
&-&h_3G_\mu\bar{q}\gamma^\mu q - h_4 G_\mu\bar{q}\gamma^\mu\gamma^{5}q\,. \label{eq:L11}
\end{eqnarray}
Here, the $b_i$ are dimensionless couplings between $X_\mu$ and $G_\mu$. The remaining parameters are as in Eqs.~\eqref{eq:SG} and \eqref{eq:Xphi}.


\providecommand{\href}[2]{#2}\begingroup\raggedright\endgroup

\end{document}